\documentclass[conference]{IEEEtran}

\usepackage[utf8]{inputenc}
\usepackage[cmex10]{amsmath}		% Math
\usepackage{amssymb}				% Symbols of the number sets (e.g., \mathbb{N})
\usepackage{siunitx}				% Typeset SI units
\usepackage[capitalise]{cleveref}	% \cref{} -> Automatically write Fig., Eq.,
	\crefformat{equation}{(#2#1#3)}		% Redefine Equation labels to be typeset as "(1)", instead of "Eq. (1)".

\ifCLASSOPTIONcompsoc	% Better handling of citations (e.g., ranged [1-5]). RECOMMENDED BY IEEE
	\usepackage[nocompress]{cite}
\else
	\usepackage{cite}
\fi

\usepackage{url}		% Better handling of URL links

\usepackage{graphicx}				% Images

\ifCLASSOPTIONcompsoc				% Subfigures -> LOADING MACRO RECOMMENDED BY IEEE
	\usepackage[caption=false,font=normalsize,labelfont=sf,textfont=sf]{subfig}
\else
	\usepackage[caption=false,font=footnotesize]{subfig}
\fi

\usepackage[table,xcdraw]{xcolor}	% Tables
\usepackage{multirow}				% Multirow

\usepackage{kbordermatrix}

\graphicspath{{Figures/}}

\begin{document}

%%%%%%%%%%%%%%%%%%%%%%%%%%%%%%%%%%%%%%%%%%%%%
% TITLE
%%%%%%%%%%%%%%%%%%%%%%%%%%%%%%%%%%%%%%%%%%%%%
\title{Joint User Mobility and Traffic Characterization in Temporary Crowded Events}

\author{
	% Authors
	\IEEEauthorblockN{
		Adriano Valadar\IEEEauthorrefmark{1},
		Eduardo Nuno Almeida\IEEEauthorrefmark{2}\IEEEauthorrefmark{3},
		Jorge Mamede\IEEEauthorrefmark{1}\IEEEauthorrefmark{2}
	}
	% Affilitations
	\IEEEauthorblockA{
		\IEEEauthorrefmark{1}ISEP, IPP, Portugal, Emails: \{1110461, jbm\}@isep.ipp.pt\\
		\IEEEauthorrefmark{2}INESC TEC, Portugal, Emails: eduardo.n.almeida@inesctec.pt \\
		\IEEEauthorrefmark{3}Faculdade de Engenharia, Universidade do Porto, Portugal \\
		%Campus da FEUP, Rua Dr. Roberto Frias, 4200-465 Porto, Portugal \\
	}
}

% Use for special paper notices

% Make the title area
\maketitle

%%%%%%%%%%%%%%%%%%%%%%%%%%%%%%%%%%%%%%%%%%%%%%%%%%%%%%%%%%%%%%%%%
% ABSTRACT AND KEYWORDS
%%%%%%%%%%%%%%%%%%%%%%%%%%%%%%%%%%%%%%%%%%%%%%%%%%%%%%%%%%%%%%%%%
\begin{abstract}

In TCEs (Temporary Crowded Events), for example, music festivals, users are faced with problems accessing the Internet. TCEs are limited time events with a high concentration of people moving within the event enclosure while accessing the Internet. Unlike other events where the user locations are constant and known at the start (e.g. stadiums), the traffic generation and the user movement in TCEs is variable and influenced by the dynamics of the event. The movement of users can lead to overloads in APs (Access Point) in case they are fixed. In order to minimize this phenomenon, new techniques have been explored that resort to the adjustable positioning of APs integrated into UAVs (Unmanned Aerial Vehicles). In these scenarios, the dynamic of the location of the APs requires that tools of prediction of the users movements and, in turn, of the sources of traffic, gain particular expression when being related to the algorithms of positioning of the referred APs. In order to allow the development and analysis of new network planning solutions for TCEs, it is necessary to recreate these scenarios in simulation, which, in turn, requires a detailed characterization of this kind of events. This article aims to characterize and model the mobility and traffic generated by users in TCEs. This characterization will enable the development of new statistical models of traffic generation and user mobility in TCEs.

\end{abstract}

\begin{IEEEkeywords}
TCEs, UAVs, Users, Movement, Traffic.
\end{IEEEkeywords}

%%%%%%%%%%%%%%%%%%%%%%%%%%%%%%%%%%%%%%%%%%%%%%%%%%%%%%%%%%%%%%%%%
% INTRODUCTION
%%%%%%%%%%%%%%%%%%%%%%%%%%%%%%%%%%%%%%%%%%%%%%%%%%%%%%%%%%%%%%%%%
\section{Introduction} \label{Introduction-Section}

\IEEEPARstart{T}{he} TCEs, such as music festivals, are events characterized by having an enclosure that is subdivided into recreational areas, such as the stage of performances, the drinks zone, the area of offer of gifts, among others. They are also characterized by the universe of people (users) that participate in the festival because each user can move along the course of the event spontaneously and unpredictability, depending on the activity they want to perform. There is another variable that characterizes these events, that is, how much a user uses the Internet, i.e. the traffic generated by each person through his smartphone. The users are faced with problems accessing the Internet because of the unpredictability of their behaviour \cite{NatalieDuffield}. These problems are due not only to the type of event, which is characterized by having traffic generation and user movement variable and dependent on the dynamics of the event, but also to the network planning that is carried out. Network planning usually considers that the network devices are fixed. Recently new techniques were developed that assume that these devices move, thus making interesting and advantageous the relationship between the movement of these devices and the movement of users.

These events are a reality on a world scale, with variety and stocking increasing exponentially. There are several cases where the tickets for the same ones run out in a short time and, in situations where entry is free, there are also cases of overcrowding in the event zone. In the case of Portugal, in the year 2017 there were 272 music festivals \cite{RicardoBramao2018}, with a growth in the number of festivals in the order of 9\% when compared to 2016.

In addition to the growth of music festivals, there is also the market for smartphones and the use of them in festivals for a wide range of purposes, from a simple call to real time video streaming. In the case of the MEO Sudoeste festival, Altice revealed the details of the 2018 edition of this festival through a press release \cite{Altice}. This announcement reveals that there has been a reinforcement at the network level, by adding 138 km of fiber and 70 new 2G, 3G and 4G cells, to ensure the full operation of the various MEO networks in the site. Of a universe of 147 thousand users, the numbers of this event are the following: 11 terabytes of WiFi data; 2.6 million unique WiFi sessions; 9089 Gigabytes of mobile data; 428 thousand telephone calls; 265 thousand SMS (Short Message Service).

The high Internet values used in this example reflects the level of the operators planning required for these events. The need to characterize, model and predict jointly the mobility and the traffic generated by the users in TCEs, motivated the elaboration of this article. Through these results it will be possible for the network devices to move according to the forecast made.

As such, a tool was developed that jointly characterizes and models the mobility and traffic generated by users, based on their position and movement, enables to estimate users position in the moments following, as well as, to quantify the number of users and the amount of traffic generated in each of these areas. This tool can later be used to simulate the movement and traffic generation of the users in TCEs, which, in turn, will allow to evaluate new solutions of network planning for this type of events. To the best of our knowledge, this is the first time that an application is developed that considers the mobility and the traffic together.

%The rest of this paper is organized as follows. \cref{RelatedWork-Section} presents some references to related works. \cref{SystemModel-Section} presents the system model of the developed tool. \cref{Algorithm-Section} describes the modeling process and the developed algorithm. \cref{Results-Section} presents the simulation results. Finally, \cref{Conclusions-Section} concludes the paper and points out the future work.

%%%%%%%%%%%%%%%%%%%%%%%%%%%%%%%%%%%%%%%%%%%%%%%%%%%%%%%%%%%%%%%%%
% RELATED WORK
%%%%%%%%%%%%%%%%%%%%%%%%%%%%%%%%%%%%%%%%%%%%%%%%%%%%%%%%%%%%%%%%%
\section{Related Work} \label{RelatedWork-Section}

In \cite{Shafiq}, the authors analyse the network failures in TCEs. Their focus was to show that users movements and placement changes, the behavior of each user and the type of applications used result in a significant degradation of network performance. This article is interesting for this work, since it proves that there are network failures in this type of events. However, no solution is presented.

In \cite{Wang2015}, data from various 3G and LTE (Long Term Evolution) cell towers of metropolitan areas is extracted and analyzed. In order to organize the data, a clustering technique, called \emph{hierarchical clustering} is used, which consists of grouping the data in pairs, these pairs being once again grouped, and so on, until it reaches the number of wanted clusters \cite{Zhao2002}. This technique is used in order to identify or define areas where users have a characteristic behavior in the generation of traffic. The main goal of that work was to create a model that combines time, location and frequency information to analyze the traffic patterns of thousands of cell towers.

In the case of article \cite{Shafiq2011}, the aim of the authors is to understand the dynamics of Internet traffic in large cellular networks, which, according to the authors, is useful for network design, problem solving, performance evaluation and optimization of the network. The data used for this study was collected from a telecommunications operator. Data corresponds to mobile traffic for a week. The method \emph{K Means Clustering} is used to categorize the various types of devices used. A traffic prediction model based on Markov chains is also created. This model, called Markov Model, has a state transition matrix defined by a set of properties, such as: number of rows equal to the number of columns and the sum of each row is always equal to 1. This model has the property that, while in a state, it is only possible to remain in the same state or transit to a contiguous state.

The articles \cite{Wang2015} and \cite{Shafiq2011} are close to the desired solution, since in both cases there is a traffic characterization and in the second case there is a traffic prediction model. However, in these cases the APs are fixed, which, in case of TCE events, is disadvantageous compared to the use of UAVs. This is due to the fact that there may be obstacles in the event venue and interference with the crowd's field of vision. These studies are also distant from the intended since they deal with cases of cellular networks, and, in our case, a Wi-Fi based network solution is desired.

There are also several other studies on this topic, where \emph{clustering} is a technique often used \cite{Tutschku1998} \cite{Ghosh2011} \cite{Kodinariya2013}. The technique \emph {K Means} is used in \cite{Kodinariya2013}.

%%%%%%%%%%%%%%%%%%%%%%%%%%%%%%%%%%%%%%%%%%%%%%%%%%%%%%%%%%%%%%%%%
% SYSTEM MODEL
%%%%%%%%%%%%%%%%%%%%%%%%%%%%%%%%%%%%%%%%%%%%%%%%%%%%%%%%%%%%%%%%%
\section{System Model} \label{SystemModel-Section}

\begin{figure}
	\centering
	\includegraphics[width=0.9\linewidth]{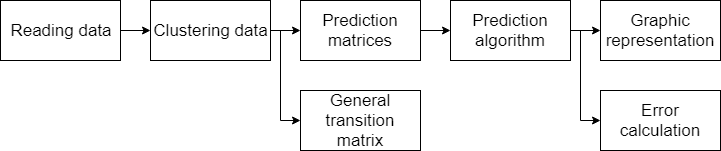}
	\caption{Stages of the system}
	\label{SystemModel-Figure: Architecture}
\end{figure}

The several stages of the tool developed for analyzing and predicting the mobility of users and traffic in each zone, are illustrated in \cref {SystemModel-Figure: Architecture}. The data readout includes the following parameters: user positions and average traffic generated by each user, boundaries of the enclosure, sliding window size, number of groups required and number of users whose graph is to be viewed. User positions can be continuous or discrete. User position data refer to different time instants, having regular intervals between them. The first instant corresponds to the initial instant 0 and from there the following instants correspond to T, 2T, until the final instant, nT, where n refers to the total of instants of the context. Both traffic and distances are considered adimensional. Taking as an example the case of the boundaries of the enclosure, the system will consider rectangular enclosures where the real distances can be calculated by the Equation \cref{eq: coordinates}, where $D$ corresponds to the distances considered by the system, $i$ to the index and $d$ to the real distances (calculated with euclidean distances), for example the coordinates (1,1) and (1,2) of the system, considering 2 m of index, have 1 of distance with respect to the system and 2 m of real distance. 

\begin {equation} \label {eq: coordinates}
d = i * D
\end {equation}

Clustering is obtained through the K Mean technique. Several matrices will be created: one that includes all users and all instants (general transition matrix), another that includes all users but considers only a range of instants of time (this interval corresponds to the size of the sliding window), and a matrix for each user considering only a set of time instants (per-user prediction matrix). In order to be able to qualitatively distinguish the actual values from the forecast values, an associated error is calculated. The graphical representation is used to compare the actual data and the predicted data for both users and the various clusters. The per-user graphs refer to their mobility, while cluster graphs are relative to the number of users (corresponding to the sum of users at a given time) and the traffic generated in that zone (corresponding to the sum of traffic generated by each user). In case of traffic generated by cluster, the traffic generated by each user is considered constant (average traffic) and analyzed as the sum of the traffic generated by the users in that cluster. Only user mobility data is predicted, and this data is used for calculations of each cluster. The mobility data also generate histograms related to the error associated with the prediction.

%%%%%%%%%%%%%%%%%%%%%%%%%%%%%%%%%%%%%%%%%%%%%%%%%%%%%%%%%%%%%%%%%
% JOINT MOBILITY AND TRAFFIC MODELLING
%%%%%%%%%%%%%%%%%%%%%%%%%%%%%%%%%%%%%%%%%%%%%%%%%%%%%%%%%%%%%%%%%
\section{Joint Mobility and Traffic Modelling} \label{Algorithm-Section}

First of all, in order to make the situation simpler and get better results with the algorithm used, it is important to use a clustering technique to group data. In this case K Means Clustering technique was chosen, and it is used within a limited area, which corresponds to the precinct. K Means groups the mobility data, and if at any moment any user data comes from outside the precinct, this data will be considered as belonging to an extra cluster. The extra cluster is generated with another instance of K Means that only considers the data outside the precinct. The clustering process is only used once and considers all time instants, so the clusters are fixed through time.
After data clustering, mobility is modeled. For this, transition matrices are used. Several types of matrices were used.
The general transition matrix is calculated taking into account the data of all the users. This matrix is square of variable size depending on the number of clusters. Each row of this array has a sum of 1. It includes all positions of all users at all times. This array is calculated based on the clusters of each user. Each line represents the cluster where the user is in and each column represents the cluster to which the user will move, that is, each value of matrix M (i, j) represents the probability transition from cluster to cluster. For example, if there are 3 clusters (0 to 2) and a user moving from cluster 0 to cluster 2, the array will be updated in row 0 and column 2. At the end of the process, each value is divided by the sum of each line of the matrix, thus giving rise to transition probabilities. An example of an array (where the cluster is called c) is as follows:
\[\renewcommand{\arraystretch}{1.35}
\kbordermatrix{
 &c_0&c_1&c_2\\
c_0&0.25&0.25&0.5\\
c_1&0.45&0.25&0.3\\
c_2&0.5&0.45&0.05\\}
\] 

Using the first line of the matrix as a reference, it is first noted that, for example, there have been two users moving from cluster 0 to cluster 2, a user moving from the cluster 0 to cluster 1 and another user who was in cluster 0 and stays there. The first line of the matrix is represented by: [1, 1, 2]. Then, the matrix is divided by the sum of the line (which is equal to 4). The result of this line is: [0.25, 0.25, 0.5]. This matrix is not used for forecasting purposes. It is used only to enable the study of the general behavior of users at any instant and to be possible to compare with the forecast values obtained.

The matrices used for forecasting purposes are similar to the general transition matrix, the difference is that these matrices consider only a sliding window size data rather than considering all instants. These matrices are constantly updated as the event progresses. For example, if the size of the sliding window is 3, it will use the values of instants 0, 1 and 2 to learn and to predict the cluster where the user will be at instant 3. To predict the value of instant 4, this matrix is updated using the values of instants 1, 2 and 3. In the last case, the value corresponding to instant 3 refers to the cluster where the user really was, not considering the prediction value that was calculated by the algorithm in the last step. This is because if the prediction algorithm fails at an instant, this value would not only affect the quality of the algorithm at that time, but would also generate problems in the following values. There are two types of forecast matrices: one general and one per user. The difference is that the general matrix uses the values of all users for the desired sliding window, whereas the per-user matrix uses the values of only one user for the desired window.

The algorithm uses the last position predicted to have reference of the cluster in which the user is, except when the user makes the forecast for the first time. In that case the algorithm will use a value that actually happened. The matrix line for the cluster number that the user is in is used to know the transition probabilities of that cluster. A random number between 0 and 1 is then generated, compared to the values of the line in question and then the cluster to which the user is expected to go is decided. This decision is made by taking the probabilities of each line as intervals. Using the example of the user that is in cluster 0, and having a generated random number equal to 0.49, this value will be compared with the intervals [0;0.25]  (range referring to the user staying in cluster 0), ]0.25; 0.5] (range referring to the user moving to cluster 1), and ]0.5;1] (range referring to the user moving to cluster 2). Thus, in this case, following the logic of the algorithm, the user will move to cluster 1.

\begin{figure}
    \centering
    \includegraphics[width=0.7\linewidth]{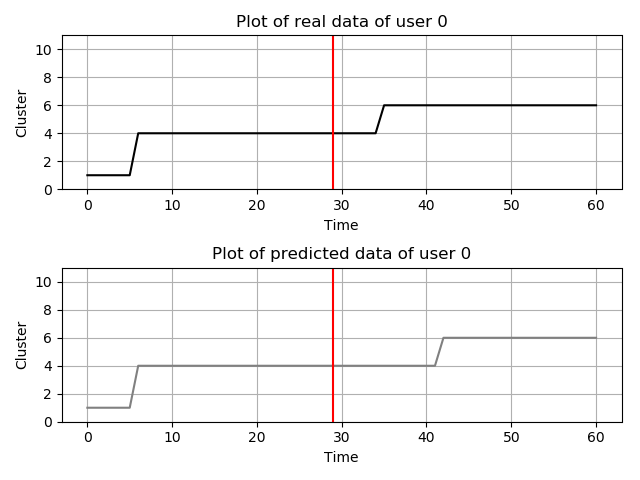}
    \caption{Example of mobility plots of an user (real data versus predicted data)}
    \label{JointUser-Figure: user_plot}
\end{figure}

It is important to note that each time instant corresponds to a data reading moment, which may have any time interval depending on the context. The actual data graph (corresponding to the graphical representation of the clusters of each user and the user mobility prediction data) is then generated are shown in \cref{JointUser-Figure: user_plot}. The vertical line corresponds to the last learning value in the prediction case, and up to this value the two representations are identical. In the case shown in the Fig. \ref{JointUser-Figure: user_plot}, there are 40 instants (from 0, initial state, to 39, final state) with instant 20 being the first to be predicted.

Through this data, plots are also generated in relation to the existing clusters: plot related to the number of users and the traffic generated by the users in a certain instant. These plots are obtained based on predicted mobility values. Regarding the graph of the number of users per zone, the sum of the users that are at each moment in each cluster is made, through the actual data and the forecast data, thus giving a perspective at the level of each cluster. The traffic generated in each moment is also calculated, considering the values of the average traffic of the users that are in each cluster. For example, if only 2 users are in cluster 0 at a given time and have traffic values of, for instance, 1 and 2 Mbit/s, then there are two users in that cluster and the traffic is 3 Mbit/s. The error associated to the mobility prediction is calculated to each user and is represented in Equation \cref{eq:error}, where $t$: Forecast time instant; $E(t)$: Forecast error at time $t$; \( C_{\text{r}}\)($t$): represents the coordinates of the centroid of the real cluster to which the user moved at time $t$; \( C_{\text{p}}\)($t$): represents the coordinates of the centroid of the predicted cluster; \( p_{\text{max}}\): coordinates of the maximum point in the $xy$ axis (including points outside the precinct if there are any); \( p_{\text{min}}\): coordinates of the minimum point in the $xy$ axis (including points outside the precinct if there are any).

\begin{equation}\label{eq:error}
    	E(t)=\frac{|C_r(t) - C_p(t)|}{p_{\text{max}} - p_{\text{min}}}
\end{equation}

%%%%%%%%%%%%%%%%%%%%%%%%%%%%%%%%%%%%%%%%%%%%%%%%%%%%%%%%%%%%%%%%%
% SIMULATION RESULTS
%%%%%%%%%%%%%%%%%%%%%%%%%%%%%%%%%%%%%%%%%%%%%%%%%%%%%%%%%%%%%%%%%
\section{Simulation Results} \label{Results-Section}

\begin{figure}
    \centering
	\includegraphics[width=0.7\linewidth]{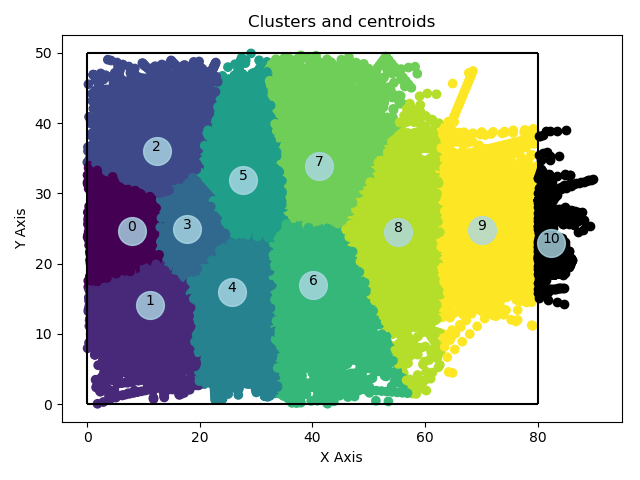}
    \caption{Simulation Plot}
    \label{Results-Figure: precinct}
\end{figure}

\begin{figure}
    \centering
	\includegraphics[width=0.7\linewidth]{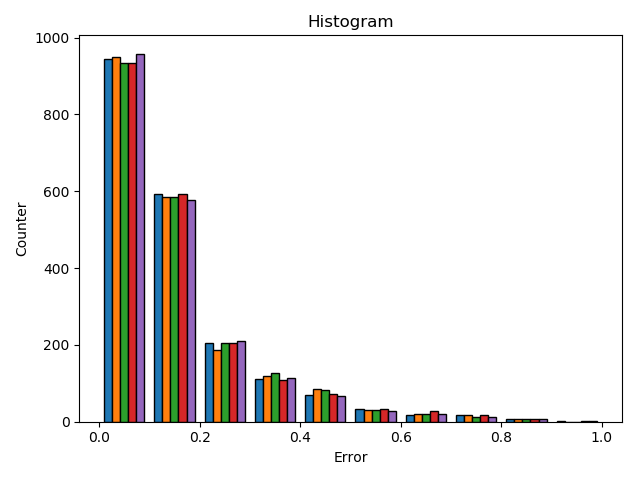}
    \caption{Histogram of simulations errors}
    \label{Results-Figure: histogram}
\end{figure}

The simulation of our scenario mobility data was obtained by the use of BonnMotion - a mobility scenario generation and analysis tool\cite{BonnMotion}, more accurately the Random Way Point algorithm. The simulation have the following properties: i) The precinct has an area of 50*80 meters, and there is an area of 50*10 meters outside the precinct (although to the side of the application this is considered adimensional); ii) There are 2000 users; iii) The duration of the event is 5 hours, the data is read in intervals of 5 minutes, so there are 60 time instants where the data is read; iv) The stage is in the middle left of the precinct; v) At the top and bottom of the precinct there are the food and drinks areas, toilets and a Ferris wheel; vi) The entrance and the exit area is done through the right limit of the precinct; vii) The users can move in a velocity between 0 and 0.08 m/s and there is a bigger probability that they will go to the stage area comparing to the other areas; viii) $\frac{1}{3}$ has a medium traffic of 0 Mb/s, $\frac{1}{3}$ of 10 Mbit/s and the other $\frac{1}{3}$ has a medium traffic of 10 Mbit/s. The plot that considers all places where the users were considering all instants of time readings can be viewed in Fig. \ref{Results-Figure: precinct}. This simulation has been run for 5 times and a histogram of the error of the mobility data was generated (shown in Fig. \ref{Results-Figure: histogram}). In this histogram each run is represented by a different colour. The achieved error is about 10\%. It is evident that the prediction of the algorithm is reasonably accurate.

%%%%%%%%%%%%%%%%%%%%%%%%%%%%%%%%%%%%%%%%%%%%%%%%%%%%%%%%%%%%%%%%%
% CONCLUSIONS
%%%%%%%%%%%%%%%%%%%%%%%%%%%%%%%%%%%%%%%%%%%%%%%%%%%%%%%%%%%%%%%%%
\section{Conclusions} \label{Conclusions-Section}

This article presented a algorithm that considers mobility and traffic jointly, that has an satisfactory error associated. Although the satisfying results, there is more work to do, such as creating an algorithm that evaluates the best size of the sliding window, considering the traffic as variable, and an evaluation of traffic equal to what has been done with the mobility. It is also future work to consider running K Means several times according to the size of the sliding window.

%%%%%%%%%%%%%%%%%%%%%%%%%%%%%%%%%%%%%%%%%%%%%%%%%%%%%%%%%%%%%%%%%
% REFERENCES
%%%%%%%%%%%%%%%%%%%%%%%%%%%%%%%%%%%%%%%%%%%%%%%%%%%%%%%%%%%%%%%%%

% trigger a \newpage just before the given reference number - used to balance the columns on the last page. adjust value as needed - may need to be readjusted if the document is modified later
%%%\IEEEtriggeratref{8}

\bibliographystyle{IEEEtran}
\bibliography{IEEEabrv,References}

\end{document}